\documentclass{article}
\usepackage{eucal}
\usepackage{amssymb}
\usepackage{eufrak}
\usepackage{amsmath}
\newtheorem{tetel}{Theorem}[section]
\newtheorem{lemma}[tetel]{Lemma}
\newtheorem{definicio}[tetel]{Definition}
\newtheorem{konvencio}[tetel]{Convention}
\title{On an implementation of the Solovay-Kitaev algorithm}
\author{Attila B. Nagy\thanks{
Computer and Automation Research Institute
of the Hungarian Academy of Sciences,
Kende u. 13-17, H-1111 Budapest, Hungary.
E-mail:
nagy@math.bme.hu
Research partially supported by the IST-FET program RESQ IST-2001-37559 of the EC.}}

\begin{document}
\maketitle

\begin{abstract}
In quantum computation we are given a finite set of gates and we have to perform a desired operation as a product of them.
The corresponding computational problem is approximating an arbitrary unitary as a product 
in a topological generating set of $SU(d)$. 
The problem is known to be solvable in time $\mbox{polylog}(1/\epsilon)$ with product length $\mbox{polylog}(1/\epsilon)$,
where the implicit constants depend on the given generators.
The existing algorithms solve the problem but they need a very slow and space consuming preparatory stage.
This stage runs in time exponential in $d^2$ and requires memory of size exponential in $d^2$.
In this paper we present methods which make the implementation of the existing algorithms easier.
We present heuristic methods which make a time-length trade-off in the preparatory step. 
We decrease the running time and the used memory to polynomial in $d$ but the 
length of the products approximating the desired operations will increase (by a factor which depends on $d$).
We also present a simple method which can be used for decomposing a unitary into a
product of group commutators for $2<d<256$, which is an important part of the existing algorithm.
\end{abstract}

\section{Introduction}
The Solovay-Kitaev theorem (\cite{cgc}, Section 8) asserts that if a set $\mathsf{G}$ in $SU(d)$ (with some simple properties) generates a dense subgroup
in $SU(d)$ then this set fills up $SU(d)$ quickly. It means that we can approximate an arbitrary unitary $U\in SU(d)$ with a short product of
operations from $\mathsf{G}$.
Let $\mathsf{G}=\{g_1,g_2,\dots g_n\}$ be a generating set for $SU(d)$. 
We would like to approximate an arbitrary $U\in SU(d)$ with arbitrary precision $\epsilon$,
we need a sequence of $g_{1_U}, g_{2_U}, \dots g_{m_U}$ ($g_{i_U}\in \mathsf{G}$) which satisfy the inequality $||U-\prod_{i=1}^{m} g_{i_U}||\le\epsilon$.
In \cite{cgc} we can find an algorithm which produce this product by constructing recursive coverings of the neighborhoods of the identity.
In \cite{solkit} we can find a nice and transparent version of the Solovay-Kitaev algorithm. 
These methods require a preparatory stage (which must be performed only once for a fixed generating set $\mathsf{G}$) 
which generates a set $\Gamma$ of products from $\mathsf{G}$ up to a fixed length, which depends on $d$. This step needs huge computational efforts
for higher $d$.
The set $\Gamma$ gives an initial $\epsilon_0$-covering of $SU(d)$ but the size of $\Gamma$ is exponential in $d^2$ and hence the time of generating $\Gamma$ is
exponential in $d^2$ too. It needs a huge computational effort for $d>2$, that is in the case of more than a single qubit.\\
In \cite{solkit} the initial covering needs $\epsilon_0\le 1/(8\sqrt{d}(d-1)/2)$. In \cite{cgc} the algorithm uses a universal $\epsilon_0$
which is a power of $1/20$. 

In this paper we suggest a data structure and heuristic methods for producing $\Gamma$ in polynomial time and space in $d$.
The payoff is that the length of a product approximating a given operation is increased by a factor
which depends on $d$. Combining this version of the preparatory stage
with the second stage of the algorithms mentioned above will give an approximation for an arbitrary unitary in $SU(d)$
as a product from $\mathsf{G}$ where the length of the product is still polylogarithmic in $1/\epsilon$ but the constant factor will be increased.
The method is scalable in the sense that the speed of the preparatory step can be chosen 
and so can the factor which increases the length of the approximating product.

The main idea is adopting the methods "Shrinking", "Telescoping" and "Zooming in" which are described in section 8 in \cite{cgc}.
In Section \ref{sec:fogalmak} we define the basic notions
which we will use in this paper. In Section \ref{sec:alg} we describe our methods and we present some computational results.
In Section \ref{sec:decompose} we extend the algorithm described in \cite{solkit} by giving an alternative method to decompose a unitary.
It is useful because the method presented in \cite{solkit} induces matrix diagonalization which appears to be difficult to implement
in some systems for symbolic computation (for instance GAP \cite{gap}). Increasing the quality of the initial covering
we can make the decomposition avoiding matrix diagonalization.

\section{The main tools}
\label{sec:fogalmak}
We review the main ideas of the Solovay Kitaev algorithm. First, we 
specify the requirements regarding the generating set $\mathsf{G}$.

\begin{konvencio} We say that $\mathsf{G}$ is a generating set for $SU(d)$ if the following hold:
\begin{enumerate}
\item $\mathsf{G}\subset SU(d)$ and $\mathsf{G}$ generates a dense subgroup in $SU(d)$.
\item If $g\in \mathsf{G}$ then $g^{-1}=g^{\dag}\in \mathsf{G}$.
\end{enumerate}
\end{konvencio}

The first requirement is natural as we have to approximate an arbitrary element of $SU(d)$ by an element from the semigroup generated by $\mathsf{G}$.
(Note that a closed subsemigroup of $SU(d)$ is a subgroup.)
The second requirement is technical. However, it is not known what 
we can say about the length of a product in an approximation if $\mathsf{G}$ does not satisfy the second
condition. However, in quantum computation the second condition is usually not really
restrictive.

For measuring the distance in $SU(d)$ we use the operator norm.\\
If $U,V\in SU(d)$ then $d(U,V)=||U-V||$ where $||U||=\sup_{|x|=1}|Ux|$. It satisfies the
following properties:
\begin{enumerate}
\item $d(U,V)\ge 0$ and $d(U,V)=0$ iff $U=V$,
\item $d(U,V)=d(V,U)$,
\item $d(U,V)\le d(U,W)+d(W,V)$,
\item $d(UW,VW)=d(WU,WV)\le d(U,V)$ (assuming $W$ is unitary),
\item if $d(U_i,U^{'}_i)\le\delta_i$ ($i=1,\ldots,l$) 
then $d(\prod_{i=1}^l U_i,\prod_{i=1}^l U^{'}_i)\le\sum_{i=1}^l \delta_i$,
assuming that $U_i,U_i'$ are unitary.
\end{enumerate}

We recall the following lemma from \cite{solkit}.

\begin{lemma}\label{lem:tavtul}
\begin{enumerate}
\item If $A,B,A^{'},B^{'}$ are unitaries such that\\$d(A,A^{'}),d(B,B^{'})\le\delta_1$, $d(I,A),d(I,B)\le\delta_2$ then
\begin{eqnarray*}
d([A,B],[A^{'},B^{'}])&\le&8\delta_1\delta_2+4\delta_1\delta_2^2+8\delta_1^2+4\delta_1^3+\delta_1^4.
\end{eqnarray*}
\item If $A,B$ are Hermitian matrices and $||A||,||B||\le \delta$ then 
\[
d(\exp(i(A+B)),\exp(iA)\exp(iB))\le \delta^2.
\]
\item If $A,B$ are Hermitians such that $||A||,||B||\le \delta$ then
\[
d(\exp(iA)\exp(iB)\exp(-iA)\exp(-iB),\exp([iA,iB]))\le 4\delta^3.
\]
\item If $A$ is Hermitian then $d(\exp(iA),I)\le ||A||$.
\end{enumerate}
\end{lemma}
The proof of the first statement can be found in \cite{solkit}. The proof of the other statements can be found in \cite{cgc} and \cite{trotter}.

We adopt the notion of nets from \cite{cgc}.

\begin{definicio}\label{def:halodef}
\begin{itemize}
\item Let $\Gamma,H\subset SU(d)$. Then $\Gamma$ is a $\delta$-net for $H$ if for all $h\in H$ there exist $\gamma\in\Gamma$ that $d(\gamma,h)\le\delta$.
\item An $(r,\delta)$-net in $SU(d)$ is a $\delta$-net for the $r$-neighborhood of the identity which denoted by $S_r$.
\item Let $\Gamma$ be a $\delta$-net for $H$. Then $\Gamma$ is $\alpha$-sparse if for all $\gamma\in\Gamma$ there exist $h\in H$ such that
$d(h,\gamma)\le\delta$ and for all $\gamma_1,\gamma_2\in\Gamma$ we have $d(\gamma_1,\gamma_2)\ge\alpha\delta$.
\item For an $(r,\delta)$-net the ratio $q=r/\delta$ is called the quality of the net.
\end{itemize}
\end{definicio}

As a part of the Solovay-Kitaev theorem in \cite{cgc} it was proved that any $\epsilon$-net in a compact semisimple Lie group generates a dense 
subgroup if $\epsilon$
is small enough. The proof is constructive in the sense that there is an algorithm to generate an
$\epsilon^{\prime}$-net from an $\epsilon$-net for an arbitrary $\epsilon^{\prime}<\epsilon$.
The algorithm presented in \cite{solkit} works in a reverse order: it recursively decomposes a unitary into a group commutator in which
the components are close to the identity and it approximates the commuting elements using an $\epsilon$-net.

We will make use of some basic properties of the nets (shown in \cite{cgc}):
\begin{lemma}\label{lemma:halotulajdonsagok}
\begin{enumerate}
\item (Telescoping)\\
Let $\Gamma_1$ be an $(r_1,\delta_1)$-net, $\Gamma_2$ be an $(r_2,\delta_2)$-net, where $\delta_1\le r_2$.\\
Then $\Gamma_1\Gamma_2=\{U_1U_2\mbox{ : }U_1\in\Gamma_1\mbox{, }U_2\in\Gamma_2\}$ is an $(r_1,\delta_2)$-net.
\item (Zooming in)\\
Let $\Gamma_0,\Gamma_1,\dots \Gamma_n\subset SU(d)$ be nets, where $\Gamma_i$ is an $(r_i,\delta_i)$-net and $\delta_i\le r_{i+1}$.
If $V\in S_{r_0}$ then $V$ can be approximated by $U=U_0U_1\cdots U_n$ where $U_i\in\Gamma_i$ and $d(U,V)\le \delta_n$.
\end{enumerate}
\end{lemma}

The main problem is that an $(r,r/q)$-net has at least $q^{O(d^2)}$ points. It follows from the fact
that in $SU(d)$ the volume of a sphere with diameter $\delta$ is $O(\delta^{d^2-1})$.
But $q^{O(d^2)}$ points are enough for an $(r,r/q)$-net, it follows from the fact that one can
make an $(r,r/q)$-net in $\mathbf{C}^{d^2}$ with $q^{O(d^2)}$ points.
For the purposes of the second stage an initial covering of $SU(d)$
with quality $q\ge 20$ is required.
 (Note that he diameter of $SU(d)$ is 2 in the operator norm, i.e. $d(x,y)\le 2$ for all $x,y\in SU(d)$.)

\section{The preparatory stage}\label{sec:alg}

To illustrate the use of zooming in, assume that $\Gamma$ is an $\epsilon$-net for the entire $SU(d)$. 
Thus $\Gamma$ has $1/\epsilon^{O(d^2)}$ elements.

We would like to show that the number of matrices in $\Gamma$ can be decreased, but the
length of products which produce the elements of $\Gamma$ will increase.

We define a sequence of nets $\Gamma_i$ 
($i=0,1\dots \lceil(1+\log(1/\epsilon))/\log(q)\rceil$, where $1<q<1/\epsilon$) 
by 
$$\Gamma_i=\{\gamma\in\Gamma\mid 2/q^i\le d(\gamma,I) \le 2/q^{i+1}\}\cup\{I\}.$$
Each $\Gamma_i$ will be
an $2/q^{i+1}$-net for $S_{2/q^i}\setminus S_{2/q^{i+1}}$, so each $\Gamma_i$ is an $(r_i,r_i/q)$-net where $r_0=2$, and
$r_{i+1}=r_i/q$. Then sparsening these sets we obtain $(r_i,r_i/q)$-nets where each $\Gamma_i$ has $q^{O(d^2)}$ elements.

Using Zooming in (defined in Lemma \ref{lemma:halotulajdonsagok}) we get an $\epsilon$-net as a chain of $(r_i,r_i/q)$-nets
$\Gamma_i$ ($i=0\dots k=(1+\log(1/\epsilon))/\log(q)$) with "poor" quality $q$.
In this case the length of a product in $\Gamma_i$ is not more than the maximal length of a product in the
$\epsilon$-net. So the length of a product using the Telescoping structure instead of the $\epsilon$-net $\Gamma$
will be increased by a factor
$k=(1+\log(1/\epsilon))/\log(q)$.
The cardinality of $\bigcup\Gamma_i=k\cdot q^{O(d^2)}$.
With an appropriate choice of $q$ (i.e. $q=\sqrt[d^2]{2}$) we can ensure that $|\bigcup\Gamma_i|$ is polynomial
in $d$, however in this case the length of the products will increase.

Computational experience shows that it is difficult to construct nets with "very poor" quality, i.e. $q\le \sqrt[m]{2}$ where $m>>d^2$.\\

The most important problem is constructing such a sequence of nets quickly.
The only known accurate method for constructing a base $\epsilon$-net is to 
compute and store all products from $\mathsf{G}$ up to a fixed length.
The problem is difficult in the sense that we do not know anything about $\mathsf{G}$ 
and the problem strongly depends on the properties of $\mathsf{G}$.
In general, without any assumption about $\mathsf{G}$ it is the only known method. 
Consider the case when $d(g,I)\le \delta$ for all $g\in \mathsf{G}$,
then the distance between the identity and an $n$-length product will be at most $n\delta$. 
But computing all products and then sparsening it leads to the same problem: 
storing a huge amount of matrices.

We propose a heuristic method which speeds up the construction of a sequence of $(r_i,r_i/q)$-nets 
by increasing the length of the products in approximations
(by an additional factor). We construct the nets in parallel.\\

\textit{The main heuristic algorithm}\\

Let $\mathsf{G}$, $q$ and $\epsilon$ be fixed, where $1<q<1/\epsilon$ as above
(i.e. $q\approx \sqrt[d^2]{2}$)
and we assume that we have an initial $(2,2/q)$-net $\Gamma_0$.

Let $k=\lceil (1+\log(1/\epsilon))/\log(q) \rceil$, and let $\Gamma_1,\Gamma_2,\dots,\Gamma_k$ be empty sets.
At the end of the procedure each $\Gamma_i$ will be hopefully an $(2/q^i,2/q^{i+1})$-net.
At first for all $g\in \mathsf{G}$ let $\Gamma_i=\Gamma_i\cup \{g\}$ iff $2/q^{i}\ge d(g,I)\ge 2/q^{i+1}$ ($i=0\dots k$).

In each step we increase the cardinality of the set $\bigcup_{i=0}^k \Gamma_i$. If we cannot increase this
cardinality then the algorithm terminates. The set $\Gamma_i$ remains a subset of an $(2/q^i,2/q^{i+1})$-net after each step,
and we assume that $\Gamma_i$s are sparse (Definition \ref{def:halodef}), 
so each $\Gamma_i$ has at most $q^{O(d^2)}$ elements after each step.

In each step of the algorithm we calculate the products $H=\mathsf{G}\cdot \bigcup\limits_{i=0}^k \Gamma_i$.
It is easy to check that
\[|H|\le |\mathsf{G}|\cdot \left(\sum\limits_{i=0}^k |\Gamma_i|\right)\approx |\mathsf{G}|\cdot k\cdot q^{O(d^2)}.\]

Let $h\in H$ (where $h=g\gamma$, $g\in \mathsf{G}$, $\gamma\in\bigcup\limits_{i=0}^k \Gamma_i$) and let $\delta=d(I,h)$.
We check the following properties for all elements of $H$:
if there is an index $i$ with $2/q^{i}\ge \delta\ge 2/q^{i+1}$ then check if there is an element $\gamma_i\in \Gamma_i$ with
$d(\gamma_i,h)\le 2/q^{i+1}$. If there is no such $\gamma_i$ then $\Gamma_i=\Gamma_i\cup\{h\}$ and we continue the algorithm
with an another element of $H$.
Otherwise we divide $h$ by the element
$\gamma_i$, in this case $d(h\gamma_i^{-1},I)\le 2/q^{i+1}$
so $h\gamma_i^{-1}$ is a candidate for membership in $\Gamma_j$ for some $j>i$. We check the same property for 
$h\gamma_i^{-1}$, and so on.
If $d(h\gamma_i^{-1},I)\le\epsilon$ then we do not use this element, elsewhere we increase the cardinality of $\bigcup\Gamma_i$ by
adding $h\gamma_i^{-1}$.
We continue the method until we cannot get new elements.

This method increases the length of the products (because of dividing), but we can control this length by choosing a maximal length $L$ and we
will not use a product which has length more then $L$, where $L$ can be chosen $c\cdot \log(1/\epsilon)$ for some constant $c$.

Our computational experience shows that in many cases once the algorithm terminating,
each of the $\Gamma_i$ is a $(2/q^i,2/q^{i+1})$-net.
Of course, the correctness and the performance of the method strongly depends on the properties of $\mathsf{G}$. 
The assumption that we have an initial $(2,2/q)$-net $\Gamma_0$ is technical. If $q$ is close to 1 ($q\approx \sqrt[d^2]{2}$),
then the cardinality of $\Gamma_0$ does not
depend on $d$ and we can construct it quickly. The interpretation of this assumption is that
we need some elements which are far from the identity (the method fills up the
nets downward).

As in each step we get a new element (except when the element is closer to the identity then $\epsilon$ or the length is more then $L$)
 the running time of the method is $O(\sum |\Gamma_i|)=O((1+\log(1/\epsilon))/\log(q)\cdot|\Gamma_i|$. Using the Telescoping method
we get an $\epsilon$-net where the length of the products is less then $L\cdot (1+\log(1/\epsilon))/\log(q)$.

The algorithm will fail when $\mathsf{G}$ has some "bad" properties. For example consider the following case:
Let $\mathsf{G}_f$ be a finite matrix group such that for each $g\in \mathsf{G}$
there exist $g^\prime\in \mathsf{G}_f$ such $d(g,g^\prime)\le\epsilon$. A product of length $n$ from
$\mathsf{G}$ is at most $n\epsilon$ far from $\mathsf{G}^\prime$.

A special case is when for each $g\in \mathsf{G}$ the distance $d(g,I)\le\epsilon$. Let $U$ be an arbitrary unitary from
the group generated by $\mathsf{G}$. In this case $d(U,Ug)=d(I,g)\le\epsilon$ so we can not get new elements with the
above algorithm for arbitrary $\Gamma_i$s.

We
tested the algorithm for the usual generating set of $SU(d)$ consisting of the
Hadamard-gate, K-gate, $\pi/8$-gate and the CNOT-gate.
The quality $q$ was selected from $\sqrt[d]{2}$ to $\sqrt[d^2]{2}$ and $d$ was selected from
$\{2,4,8\}$.
We obtained that we could produce an $\epsilon$-net with this method quickly (mainly because the size of 
the data structure representing the net is not exponential in $d^2$).
With $q=\sqrt[d]{2}$ the size of $\Gamma_i$ will be $\sqrt[d]{2}^{O(d^2)}$, $k=d\cdot \log(1/\epsilon)$ and $L=O(d^2\cdot \log(1/\epsilon))$.
Using the method of Telescoping the length of a product will be at most $kL$.
With the choice $q=\sqrt[d^2]{2}$ we obtained that $|\Gamma_i|=c$ where $c$ depends on $\mathsf{G}$. 
In our test cases $5\le c\le40$.

\bigskip

\textit{A complementary method for further possible speedup}\\
Let $\Gamma$ be an $(r,r/q)$-net where $q>4$. 
This method produces a set of elements $\Gamma^{'}$ and computational results show
that $\Gamma^{'}$ is often an $(r/q,r/q^2)$-net. The length of a product in $\Gamma^{'}$ is three times the length of the products in $\Gamma$.

The method is the following,
let $H=\Gamma\cap S_{r/2}=\{U\in \Gamma: d(U,I)\le r/2\}$. If $q>4$ then $H$ is not the empty set, and so $H$ is an $(r/2,(r/2)/(q/2))$-net.
We have $HH\subset S_r$ since 
\[
d(UV,I)=||UV-I||=||(U-I)V+(V-I)||\le 
\] \[
||U-I||+||V-I||=d(U,I)+d(V,I)\le r/2+r/2=r
\]
for $U,V\in H$.

As $\Gamma$ is an $(r,r/q)$-net, if $UV\in HH$ then there exist $W\in\Gamma$ such $d(UV,W)\le r/q$ and in this case $d(UVW^{-1},I)\le r/q$.\\
Let $\Gamma^{'}=\{UVW^{-1}:U,V\in H, W\in\Gamma, d(UV,W)\le r/q\}$.
\\Then
$|\Gamma^{'}|\le|H|\cdot|H|=(q/2)^{O(d^2)}\cdot (q/2)^{O(d^2)}$. The length of a product in $\Gamma^{'}$ is at most three times as large as
the maximal length of a product in $\Gamma$. 

Obviously, $\Gamma^{'}$ will not be a net in all cases. For instance,
let $\Gamma$ be a finite subgroup of $SU(d)$ which is also a $(2,2/q)$-net, but it is not
a net with quality $q^{\prime}>q$. The method will not work for this $\Gamma$.
Computational results show that for the usual generating sets and with $q>4$ the set $\Gamma^{'}$ will be an $(r/q,r/q^2)$-net
(however, often it is much more dense than desirable).

\bigskip

For testing we used the following gates as generating set $\mathsf{G}$:
\begin{eqnarray*}
\mbox{Hadamard gate}&&\frac{1}{\sqrt{2}}\left(\begin{array}{rr}1&1\\1&-1\end{array}\right)\\
\mbox{K-gate}&&\left(\begin{array}{rr}1&0\\0&i\end{array}\right)\\
\pi\mbox{/8-gate}&&\left(\begin{array}{rr}1&0\\0&e^{i\pi/8}\end{array}\right)\\
\mbox{CNOT-gate}&&\left(\begin{array}{rrrr}1&0&0&0\\0&1&0&0\\0&0&0&1\\0&0&1&0\end{array}\right)\\
\end{eqnarray*}

\section{Decomposing a unitary}\label{sec:decompose}
In this section we extend the algorithm described in \cite{solkit}. We give a brief outline of the algorithm.
The method is recursive, for a unitary $U$ it gives
an $\epsilon_n$ approximation $U_n$ in the $n$-th iteration step. It decomposes the quotient $\Lambda=UU_n^{-1}$ into a group commutator $\Lambda=[V,W]$
where $V$ and $W$ are close to the identity and it performs the algorithm on $V$ and $W$ as well.
This algorithm needs an $\epsilon_0$ covering for $SU(d)$
and in each iteration step it gives an $\epsilon_n$ approximation for $U$ where $\epsilon_n\to 0$ as $n\to\infty$; more precisely
$\epsilon_n=c_{approx}\epsilon_{n-1}^{3/2}$ for a constant $c_{approx}$. Hence $\epsilon_n\to 0$ if $\epsilon_0< 1/c_{approx}^2$.

For a complete description of this algorithm the reader is referred to \cite{solkit}. A main step of this method is to decompose a
unitary into a group commutator. There is an efficient method for unitaries in $SU(2)$ and it has been proved that the decomposition can be made for $d>2$.
But for $d>2$ the method  described in \cite{solkit} needs the diagonalization of a unitary in order to obtain a decomposition.

We give a method for decomposing a unitary into a product of group commutators and we prove that the algorithm still remains correct
but the constant $c_{approx}$ will be increased and hence we need a better $\epsilon_0$ covering. In return, the decomposition can be made easier.

The algorithm in the $(n-1)$-th iteration step gives an approximation $U_{n-1}$ for a unitary $U$ where $d(U,U_{n-1})\le \epsilon_{n-1}$.
Let $\Lambda=UU_{n-1}^{-1}$ and $d(\Lambda,I)\le \epsilon_{n-1}$. 
We decompose $\Lambda$ into a product of group commutators $[E^{(1)},E^{(2)}][F^{(1)},F^{(2)}]$ such that 
$d(\Lambda,[E^{(1)},E^{(2)}][F^{(1)},F^{(2)}])\le c_{gc_1}\epsilon_{n-1}^{3/2}$. The constant $c_{gc_1}$ will be
specified later. The unitaries $E^{(j)},F^{(j)}$ are close to the identity:
\begin{eqnarray*}
d(E^{(j)},I)&\le& c_{gc_2}\sqrt{\epsilon_{n-1}}\\
d(F^{(j)},I)&\le& c_{gc_2}\sqrt{\epsilon_{n-1}}
\end{eqnarray*}
for $j=1,2$ and for an another constant $c_{gc_2}$. We perform the algorithm on these elements with
$n-1$ iteration step and we get $E^{(j)}_{n-1},F^{(j)}_{n-1}$ where
\begin{eqnarray*}
d(E^{(j)},E^{(j)}_{n-1})&\le& \epsilon_{n-1}\\
d(F^{(j)},F^{(j)}_{n-1})&\le& \epsilon_{n-1}.
\end{eqnarray*}

The decomposition is based on the following lemma (which can be found in \cite{solkit})
\begin{lemma}\label{lem:felbontas}Let $H$ be a traceless off-diagonal $d$-dimensional Hermitian matrix. Then we can find Hermitian $F$ and $G$ such that:
\begin{eqnarray*}
[F,G]&=&iH,\\
||F||,||G||&\le&d^{1/4}\left(\frac{d-1}{2}\right)^{1/2}\sqrt{||H||}.
\end{eqnarray*}
\end{lemma}
\textbf{Proof:} Let $G$ be a diagonal matrix with the following entries:\\ $G_{j,j}=-(d-1)/2+(j-1)$. In this case $||G||=(d-1)/2$. Let $F$ be the following:
\begin{eqnarray*}
F_{j,k}&=&\left\{\begin{array}{ll} \frac{iH_{j,k}}{G_{k,k}-G_{j,j}}&if j\neq k\\0 & if j=k\end{array}\right.
\end{eqnarray*}
It is easy to see that $[F,G]=iH$, $||F||^2\le tr(F^2)\le tr(H^2)\le d||H||^2$, so $||F||\le\sqrt{d}||H||$. Rescaling $F$ and $G$ gives the desired
Hermitians.
From this lemma we can see that $c_{gc_2}$ will be $d^{1/4}((d-1)/2)^{1/2}$.

If $H$ is diagonal then conjugating it with the $d$-dimensional Fourier-matrix (or with a $d$-dimensional Hadamard which 
can be found in the database \cite{sloane} for all $d$ satisfying $4|d$ and $d<256$)
we get an off-diagonal matrix which can be decomposed.
In the original algorithm we have to diagonalize the Hermitian $H$.
Conjugating preserves the decomposition hence $S[A,B]S^{-1}=[SAS^{-1},SBS^{-1}]$.

We make the decomposition in the following way: $\Lambda$ is a unitary with traceless Hamiltonian $H$ and we write $H=H_o+H_d$
where $H_o$ is the off-diagonal
part of $H$ and $H_d$ is the diagonal part of $H$. Then $||H_o||,||H_d||\le\sqrt{d}||H||$, since $H_o,H_d$ are still Hermitians.

By Lemma \ref{lem:felbontas} $iH_o=[e^{(1)},e^{(2)}]$, $iH_d=[f^{(1)},f^{(2)}]$ where 
\[||e^{(j)}||,||f^{(j)}||\le d^{1/2}\left(\frac{d-1}{2}\right)^{1/2} \sqrt{||H||}.\]
Let $E^{(j)}=\exp(ie^{(j)})$ and $F^{(j)}=\exp(if^{(j)})$, then

\begin{eqnarray*}
&&d([E^{(1)}_{n-1},E^{(2)}_{n-1}][F^{(1)}_{n-1},F^{(2)}_{n-1}],\Lambda)\le\\
&\le&d([E^{(1)}_{n-1},E^{(2)}_{n-1}][F^{(1)}_{n-1},F^{(2)}_{n-1}],[E^{(1)},E^{(2)}][F^{(1)},F^{(2)}])+\\
&+&d([E^{(1)},E^{(2)}][F^{(1)},F^{(2)}],\Lambda).
\end{eqnarray*}
By Lemma \ref{lem:tavtul} and the properties of the distance function,
\begin{eqnarray*}
&&d([E^{(1)}_{n-1},E^{(2)}_{n-1}][F^{(1)}_{n-1},F^{(2)}_{n-1}],[E^{(1)},E^{(2)}][F^{(1)},F^{(2)}])\le\\
&\le&d([E^{(1)}_{n-1},E^{(2)}_{n-1}],[E^{(1)},E^{(2)}])+d([F^{(1)}_{n-1},F^{(2)}_{n-1}],[F^{(1)},F^{(2)}])\le\\
&\le&16\epsilon_{n-1} (d(d-1)/2)^{1/2} \sqrt{\epsilon_{n-1}}.
\end{eqnarray*}

By Lemma \ref{lem:felbontas},
\begin{eqnarray*}
&&d([E^{(1)},E^{(2)}][F^{(1)},F^{(2)}],\Lambda)=\\
&=&d([\exp(ie^{(1)}),\exp(ie^{(2)})][\exp(if^{(1)}),\exp(if^{(2)})],\exp(H))=\\
&=&d([\exp(ie^{(1)}),\exp(ie^{(2)})][\exp(if^{(1)}),\exp(if^{(2)})],\exp([ie^{(1)},ie^{(2)}]+[if^{(1)},if^{(2)}]))\le\\
&\le&d(\exp([ie^{(1)},ie^{(2)}]+[if^{(1)},if^{(2)}]),\exp([ie^{(1)},ie^{(2)}])\exp([if^{(1)},if^{(2)}]))+\\
&+&d(\exp([ie^{(1)},ie^{(2)}])\exp([if^{(1)},if^{(2)}),[\exp(ie^{(1)}),\exp(ie^{(2)})][\exp(if^{(1)}),\exp(if^{(2)})]).
\end{eqnarray*}

By Lemma \ref{lem:tavtul},
\[
d(\exp([ie^{(1)},ie^{(2)}]+[if^{(1)},if^{(2)}]),\exp([ie^{(1)},ie^{(2)}])\exp([if^{(1)},if^{(2)}]))\le d||H||^2\le d\epsilon_{n-1}^2
\]
and
\[
d(\exp([ie^{(1)},ie^{(2)}])\exp([if^{(1)},if^{(2)}]),[\exp(ie^{(1)}),\exp(ie^{(2)})][\exp(if^{(1)}),\exp(if^{(2)})])\le
\]
\[
\le8(d(d-1)/2)^{3/2}\epsilon_{n-1}^{3/2}.
\]
So the constant $c_{gc_1}$ will be $8(d(d-1)/2)^{3/2}$.

We get
\[
d([E^{(1)}_{n-1},E^{(2)}_{n-1}][F^{(1)}_{n-1},F^{(2)}_{n-1}],\Lambda)\le
\]
\[
(16(d(d-1)/2)^{1/2}+d\epsilon_{n-1}^{1/2}+8(d(d-1)/2)^{3/2})\epsilon_{n-1}^{3/2}.
\]

Let $c_{approx}=16(d(d-1)/2)^{1/2}+d+8(d(d-1)/2)^{3/2}$ and we get the desired form,
$\epsilon_n\le c_{approx}\epsilon_{n-1}^{3/2}\le \epsilon_0^{(3/2)^n}$ if $\epsilon_0\le1/c_{approx}^2$.
The length of a product at the $n$-th iteration step is $l_n=8l_{n-1}$.
To approximate a unitary with error $\epsilon$ then we need at least
\[
n>\frac{\log\left(\frac{\log(\epsilon)}{\log(\epsilon_0)}\right)}{\log(3/2)}.
\]
The length of the product will be $l_08^n=O(\log(1/\epsilon))$. But with this method the initial covering must be better then in \cite{solkit}.

\end{document}